# Emittance preservation in a plasma-wakefield accelerator

**C. A. Lindstrøm[1,2,*], J. Beinortaitė[1,3], J. Björklund Svensson[1], L. Boulton[1,4,5], J. Chappell[3], S. Diederichs[1,6], B. Foster[7], J. M. Garland[1], P. González Caminal[1,6], G. Loisch[1], F. Peña[1,6], S. Schröder[1], M. Thévenet[1], S. Wesch[1], M. Wing[1,3], J. C. Wood[1], R. D'Arcy[1] & J. Osterhoff[1]**

**Radio-frequency particle accelerators are engines of discovery, powering high-energy physics and photon science, but are also large and expensive due to their limited accelerating fields. Plasma-wakefield accelerators (PWFAs) provide orders-of-magnitude stronger fields in the charge-density wave behind a particle bunch travelling in a plasma, promising particle accelerators of greatly reduced size and cost. However, PWFAs can easily degrade the beam quality of the bunches they accelerate. Emittance, which determines how tightly beams can be focused, is a critical beam quality in for instance colliders and free-electron lasers, but is particularly prone to degradation. We demonstrate, for the first time, emittance preservation in a high-gradient and high-efficiency PWFA while simultaneously preserving charge and energy spread. This establishes that PWFAs can accelerate without degradation—essential for energy boosters in photon science and multistage facilities for compact high-energy particle colliders.**

In a conventional radio-frequency (RF) particle accelerator, the accelerating field is limited to approximately 100 MV/m by breakdowns in the metallic accelerator cavities. Consequently, X-ray free-electron lasers[1,2] (FELs), used in photon-science research with energy of order 10 GeV, are long and expensive. This is even more so for linear electron-positron colliders at the TeV scale[3,4]. By exchanging the accelerating medium from a metal to a plasma, which is not limited by breakdown effects, plasma-based accelerators can provide accelerating fields as high as 100 GV/m, 1000 times larger than RF accelerators. In principle, this promises to make accelerators significantly shorter and cheaper.

Plasma-based acceleration[5–8] can occur when an intense laser pulse or charged-particle beam (known as a *driver*) traverses a plasma, expelling plasma electrons in its path and driving a charge-density wave behind it—a so-called *plasma wake*. The resulting separation of electrons and ions creates strong electromagnetic fields, or *plasma wakefields*, which can be used both to accelerate and focus a trailing particle bunch. In beam-driven plasma-wakefield accelerators (PWFAs), experiments have already demonstrated large energy gain[9,10], high energy-transfer efficiency from the driver to the trailing bunch[11], acceleration across multi-metre-scale accelerator stages[12], as well as potential for high repetition rate[13].

Excellent *beam quality* is also required for many applications. This includes high charge, short bunch length, low energy spread, and low emittance—all different facets of a high charge density in phase space, also known as *beam brightness*. In particular, emittance, which determines how tightly a beam can be focused, strongly affects the performance of FELs and linear colliders. Typically, FELs demand 100-pC-scale bunches of sub-100 fs duration with 0.1% energy spreads and emittances of order 1 mm-mrad. Linear colliders, on the other hand, require nC-scale charge, sub-1% energy spread, and emittances as low as 0.01 mm-mrad. For plasma accelerators to be a compact, more affordable alternative to RF accelerators, each stage must not only accelerate with high gradient, efficiency, and repetition rate, but also preserve these beam qualities.

Recent experiments have demonstrated that both charge and energy spread can be preserved in a PWFA[14–16], and that a sufficient beam brightness can be maintained during a small energy boost while still allowing FEL gain at infrared wavelengths to occur[17]. However, preservation of emittance at the level required for scaling to large energy gain has until now not been established.

A beam's root-mean-square (rms) *normalized emittance,*[18] $\varepsilon_n$, represents the area of its rms ellipse in transverse phase space, given by $\epsilon_n^2 = \langle x^2\rangle\langle u_x^2\rangle - \langle x\, u_x\rangle^2$, where $x$ is the offset from the nominal trajectory and $u_x = p_x/m\,c$ is the transverse momentum normalized by the particle mass $m$ and the speed of light in vacuum $c$. This quantity is preserved during both acceleration and beam focusing provided the focusing field is linear (i.e., proportional to the transverse offset), as is the case in ideal quadrupole magnets. Similarly, in the uniform ion channel of a nonlinear plasma accelerator operating in the blowout regime[19,20], the focusing field is also linear, and thus the emittance of an accelerating electron bunch can, in principle, be preserved.

[1]Deutsches Elektronen-Synchrotron DESY, Hamburg, Germany. [2]Department of Physics, University of Oslo, Oslo, Norway. [3]University College London, London, UK. [4]SUPA, Department of Physics, University of Strathclyde, Glasgow, UK. [5]The Cockcroft Institute, Daresbury, UK. [6]Universität Hamburg, Hamburg, Germany. [7]John Adams Institute, Department of Physics, University of Oxford, Oxford, UK. *e-mail: carl.a.lindstroem@desy.de

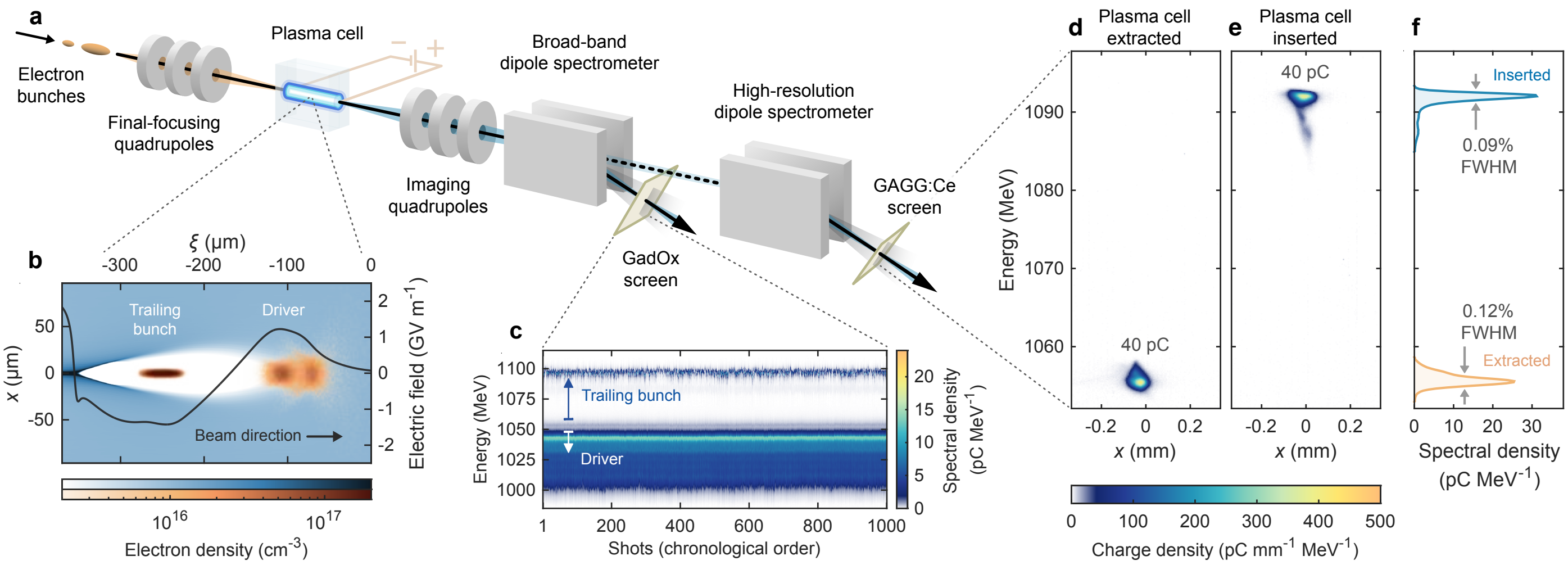

**Fig. 1 | Setup. a**, Two electron bunches were focused by quadrupole magnets into a plasma created by a high-voltage discharge, then captured and imaged with another set of quadrupoles onto one of two dipole spectrometers. **b**, A PIC simulation with plasma (blue colour scale) and beam electrons (orange colour scale) shows the leading driver bunch creating a wake in which a trailing bunch experiences GV $m^{-1}$ on-axis accelerating fields (black line) and strong transverse focusing; $x$ and $\xi = z - ct$ denote the directions perpendicular and parallel to the direction of motion, respectively. **c**, The resulting energy spectrum, measured by a broad-band spectrometer, shows that the driver loses energy (white arrow) and the trailing bunch gains energy (blue arrow), with high stability. **d–f**, Representative shots on a downstream high-resolution spectrometer show that the trailing bunch had consistent charge before (**d**) and after acceleration (**e**), and (**f**) a slightly reduced full-width-at-half-maximum (FWHM) energy spread in the accelerated spectrum (blue area) compared to the initial spectrum (orange area). All emittance measurements were performed using the high-resolution spectrometer.

However, many sources of emittance growth can complicate this picture[21]. Firstly, a bunch externally injected into a plasma-accelerator stage must be tightly focused to fit within the 10–100 μm-scale plasma cavity, and its *beta function*[22] (i.e., the Rayleigh range for a focused particle beam) must be precisely matched to the strong focusing forces therein to prevent an oscillation of the beam size[23]. Any mismatch causes phase mixing in bunches with finite energy spreads[24] and can lead to sampling of the nonlinear focusing fields near the edge of the cavity, both of which increase emittance. Similar effects occur if the bunch is transversely misaligned[25,26]. The wakefield driver can also indirectly cause emittance growth; in certain cases, particle-beam drivers can develop a hose instability[27], which leads to rapid fluctuation of the fields experienced by the trailing bunch. Additionally, if the beam driver has sufficient charge density, it can move ions towards the axis, forming an ion-density spike with highly nonlinear focusing fields[28]. Lastly, Coulomb collisions between beam and gas or plasma particles can increase the emittance through scattering[29,30]. To avoid emittance growth, all these effects must be evaluated and, if necessary, mitigated.

## Results

### Experimental setup

In this work, we demonstrate preservation of emittance in a beam-driven plasma-accelerator stage, while simultaneously preserving charge and energy spread. This was accomplished at the FLASHForward plasma-accelerator facility[31] at DESY, employing stable and high-quality beams from the FEL facility FLASH[32]. Electron bunches from a photocathode source were accelerated to 1050 MeV by superconducting RF cavities, compressed in two magnetic chicanes, and linearized in longitudinal phase space by a third-harmonic cavity. Active-feedback systems were used to stabilize the charge, energy, orbit, and bunch length. Two bunches were created in a horizontally dispersive section using a three-component mask[33]: two block collimators to remove the head and tail of the bunch, and a notch collimator to split it into a driver- and trailing-bunch pair (see the "Methods" section). Downstream quadrupole and sextupole magnets in a region of large dispersion were then adjusted to align the two bunches transversely. In the subsequent straight section, nine quadrupole magnets were used to focus the beam strongly at the entrance of a 50 mm-long capillary filled with argon gas (see Fig. 1a), around which two beam-position monitors (BPMs) measured the beam trajectory. The beam arrived 9.68 μs after a high-voltage discharge ionized the argon, resulting in a central plasma density of approximately $1.2{\times}10^{16}$ $cm^{-3}$ with upstream and downstream density ramps (see "Methods" and Supplementary Fig. 4).

The main diagnostics, downstream of the plasma-accelerator stage, were two electron energy spectrometers based on 1 m-long dipole magnets; one for broad-band spectrum measurements on a gadolinium-oxysulfide (GadOx) screen situated outside the vacuum, and another for high-resolution, energy-resolved emittance measurements on an in-vacuum cerium-doped gadolinium-aluminium-gallium-garnet (GAGG:Ce) screen. Five quadrupoles were used to capture and point-to-point image the electron beam from the plasma-cell exit (the object plane) to one of the two screens (the image plane).

### Characterization of the operating point

A multi-parameter optimization varying the incoming electron beam and the plasma density, as developed in a previous publication[14], resulted in the operating point visualized by the particle-in-cell (PIC) simulation shown in Fig. 1b (see "Methods" and Supplementary Fig. 10), which indicates a peak accelerating field of approximately 1.4 GV $m^{-1}$. The trailing bunch gained up to 40 MeV of energy per particle at an energy-transfer efficiency of around 22% (see “Methods” and Supplementary Fig. 6), measured with the broad-band spectrometer (see Fig. 1c), and had approximately 40 pC of charge both before and after acceleration, measured with the high-resolution spectrometer (see Fig. 1d–e). The reduced energy spread of the accelerated spectrum (see Fig. 1f) together with the observed high energy-transfer efficiency indicate that the wakefield was strongly beam loaded[34]. This effect is also observed in the PIC simulation, which indicates that the wakefield was under-loaded in the low-density ramp regions and over-loaded in the high-density central region, resulting in an approximately uniform wakefield when longitudinally averaged (see “Methods”). In this case, where a small low-energy distribution tail was introduced during acceleration, the energy spread is quantified using the full-width at half maximum (FWHM), as this correlates better with peak spectral density (the quantity most relevant to applications) compared to the more conventional rms (see “Methods”).

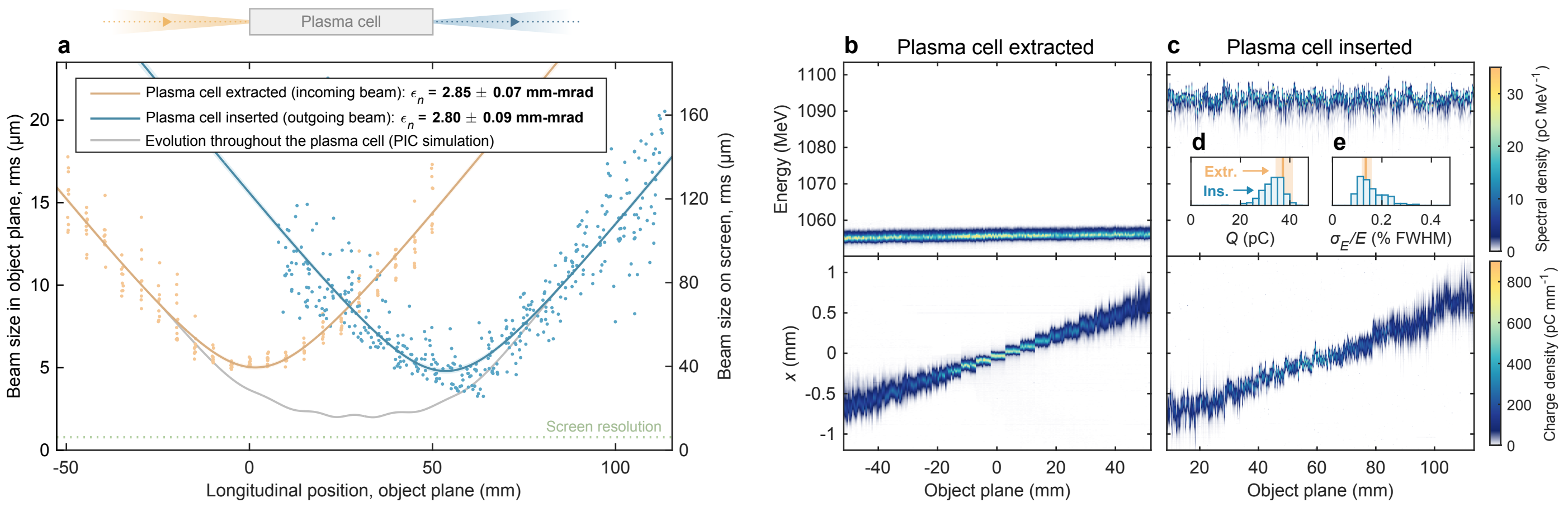

**Fig. 2 | Preservation of projected, normalized emittance. a**, The imaged beam size is shown for a range of object planes around the plasma cell, measured with the plasma cell extracted (orange points) and inserted into the beam path (blue points). The screen resolution (green dotted line) is negligible. Note that the imaged beam size does not represent the beam size as it was inside the plasma cell, but instead that of the resulting virtual waist. Fits of the virtual-waist evolution (orange and blue lines) demonstrate that the normalized emittance, $\varepsilon_n$, was preserved to within the fit error. The evolution of the beam size throughout the plasma cell is estimated using a PIC simulation (gray line). **b–c**, The measurement was performed by scanning the object plane of a point-to-point imaging spectrometer, first with the plasma cell extracted (**b**, 210 shots) and subsequently inserted (**c**, 420 shots); projections in energy and transverse position are displayed in the upper and lower panels, respectively. **d–e**, The insets show the charge, $Q$, and relative energy spread, $\sigma_E/E$, before (orange lines) and after acceleration (blue histograms).

## Preservation of emittance

Figure 2 demonstrates preservation of the projected (i.e., averaged over all energy slices), normalized emittance in the horizontal plane; starting at 2.85 ± 0.07 mm-mrad, measured with the plasma cell extracted, and ending up at 2.80 ± 0.09 mm-mrad after acceleration in the plasma. The root-mean-square (rms) horizontal beam size was measured across a range of object planes by varying the strength of the imaging quadrupoles (see “Methods” and Supplementary Fig. 7), while keeping a constant magnification as well as a constant object plane in the vertical (dispersive) plane to ensure high energy resolution (see Figs. 2b–c). This multi-shot measurement was only possible due to the high stability of the interaction (see Fig. 1c). The divergence was measured to be 0.28 mrad rms both before and after acceleration, with corresponding virtual-waist beam sizes of 5.0 and 4.7 μm rms. The screen resolution, measured to be 6.2 μm rms (see “Methods” and Supplementary Fig. 5), affected the measurement minimally, as the quadrupole imaging magnified the beam size by a factor 7.9, thereby allowing sub-μm beam features to be resolved. The preservation of emittance was achieved simultaneously with that of charge and relative energy spread: these were within the 68$^{th}$ percentile range of their initial values in 41% and 62% of all shots, respectively (see Figs. 2d–e).

## Comparison to particle-in-cell simulations

The evolution of the beam inside the plasma can be estimated via simulation (see Fig. 2a). This suggests that the trailing bunch was focused down to a beam size of less than 2 μm rms, undergoing 880° of phase advance (i.e., nearly five betatron envelope oscillations). The emittance was preserved even in the presence of a small mismatch of the beta function; the expected emittance growth after full phase mixing[24] is 10%, but this was never reached because the decoherence length for a per-mille-level energy spread would be tens of metres. Moreover, since the driver was focused 21.3 ± 0.3 mm upstream compared to the trailing bunch (due to the chromaticity of the final-focusing quadrupoles) and had a higher emittance, the transverse size of the driver was relatively large, which both suppressed the hose instability[35] and resulted in negligible motion of argon ions on the timescale of one plasma oscillation. Emittance growth from Coulomb scattering, estimated analytically from the simulation to be $1.1\times10^{-4}$ mm-mrad, was also negligible due to the small beta function inside the plasma cell[29].

## Emittance growth from misalignment

Ultimately, the main experimental challenge was to reduce misalignment and mismatching of the incoming bunch sufficiently to avoid sampling the nonlinear focusing fields in the electron sheath surrounding the plasma cavity. The emittance-preserving operating point shown in Fig. 2 was found using high-precision scans of two key parameters: the angle between the trajectories of the driver and the trailing bunch (see Fig. 3), and the longitudinal waist location of the focused trailing bunch (see Fig. 4). At each point in these scans, an object-plane scan such as that shown in Fig. 2 was performed.

Figure 3 shows the effect of misalignment on the emittance. The angle between the driver and the trailing bunch was scanned by varying the horizontal dispersion with a quadrupole magnet in the upstream dispersive section. Since the mean energy of the two bunches was slightly different (by 0.9%), this dispersion resulted in a relative misalignment (by up to $\Delta x' = \pm 1.2$ mrad). However, because the corresponding range of quadrupole strengths (±1.5%) as well as the beam size in this quadrupole were both small, the beam-waist location remained within a range of 2–7 mm downstream of that of the emittance-preserving operating point, while the waist beta function changed by less than ±25% (see "Methods" and Supplementary Fig. 2). The initial emittance were measured at optimal alignment ($\Delta x' \approx 0$), where modelling of induced intra-bunch dispersion adds up to ±6% to the uncertainty of the projected emittance for misaligned bunches (see "Methods").

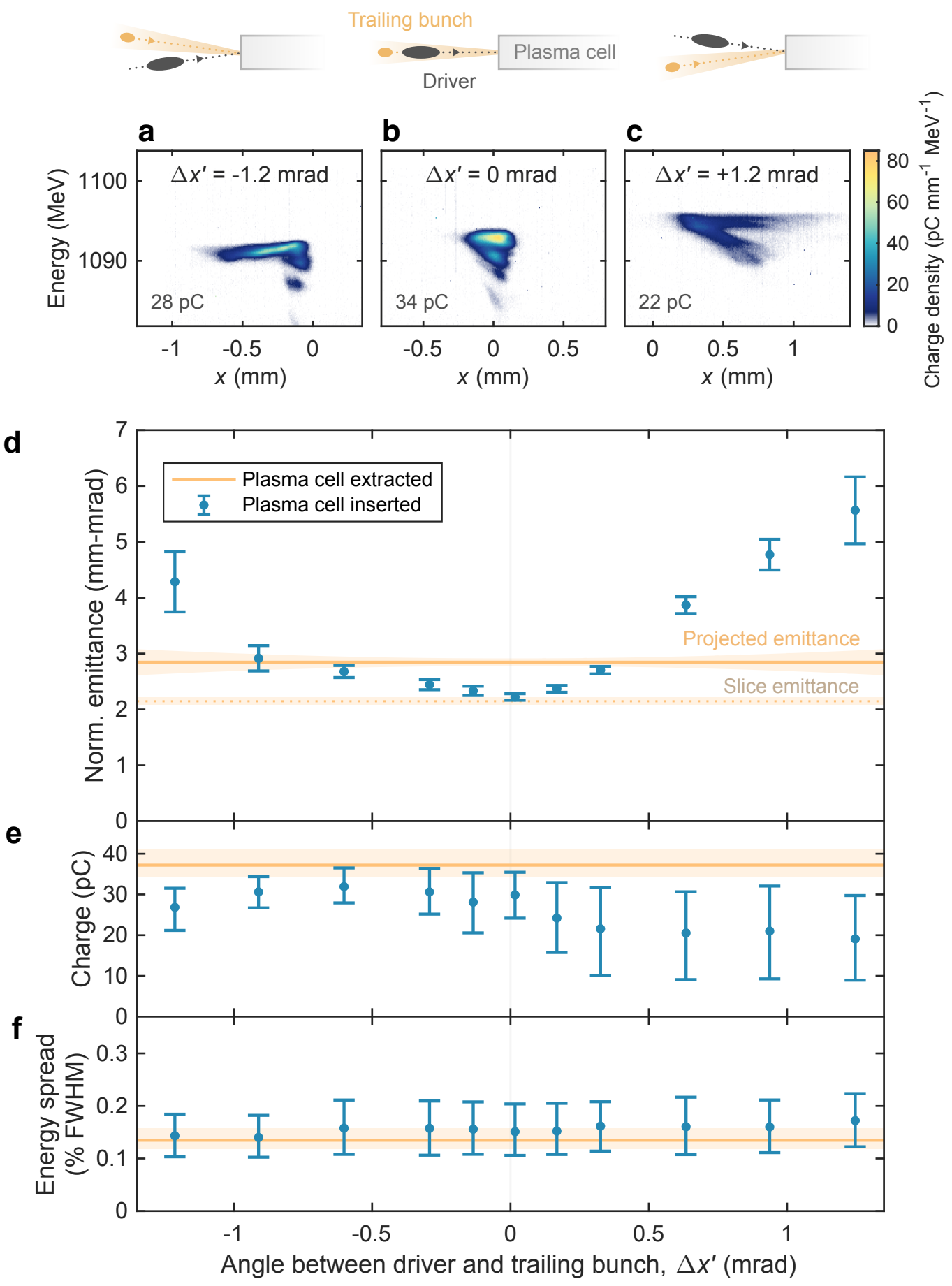

**Fig. 3 | Emittance growth due to misalignment. a–c**, Spectrometer images, captured with the plasma cell inserted and at an object plane 65 mm upstream of the plasma-cell exit, show how the accelerated trailing bunch is distorted by misalignment (**a** and **c**) compared to optimal alignment (**b**). This scan was performed at a beam-waist location 2–7 mm downstream of that of the emittance-preserving operating point (see Fig. 2), resulting in a small charge loss around the optimal alignment. **d**, Projected emittance measurements (blue error bars) are shown for a scan of angles between the driver and trailing bunches. The initial projected emittance (orange line), measured with the plasma cell extracted and at optimal alignment, where the error (light orange area) increases with misalignment to account for emittance growth from dispersion. The initial slice emittance is also shown (orange dotted line). **e**, A somewhat asymmetric charge loss is observed, likely caused by a small initial angular dispersion. **f**, The energy spread remained preserved throughout the scan. In **d**, error bars represent the best-fit value and fit error, whereas in **e–f** they represent the median and 68th percentile range of the shot distributions.

After inserting the plasma cell, optimal alignment resulted in a projected emittance somewhat lower than the initial projected emittance, close to the initial slice emittance of the central energy slice. This may be explained by an intrinsic intra-bunch dispersion within the trailing bunch (measured to be 0.1 mrad per 0.1% of energy) that exists even as the driver and trailing bunch centroids were aligned. If charge is lost from the tail of the trailing bunch during acceleration, which is consistent with observations (central error bar in Fig. 3e), this can reduce the intra-bunch dispersion and hence decrease the projected emittance. Away from optimal alignment, the emittance was observed to grow with increased misalignment. The spectrometer images in Figs. 3a and 3c, corresponding to large misalignments, show evidence of interaction with nonuniform focusing fields, deflecting the bunch tail and resulting in a higher charge loss. An asymmetry is also observed in Fig. 3d and 3e, likely caused by the intrinsic intra-bunch dispersion and the small shift in waist location and beta function across the scan.

## Emittance growth from mismatching

Using optimally aligned bunches, the matching of the beam was varied. Figure 4 shows a scan of the beam-waist location across a 33-mm range around the plasma-cell entrance, performed by fine-tuning the strength of a final-focusing quadrupole (by ±0.65%). The optics in the final-focusing section was set up such that the beam size in this quadrupole was much larger in the horizontal plane than in the vertical plane, allowing the horizontal waist location to be adjusted independently of the vertical waist location. The driver- and trailing-bunch waist locations were measured separately at each step of the scan using a two-BPM measurement technique[36] where the distribution of the orbit jitter serves as a proxy for the beam (see "Methods"). This measurement also indicated that the relative separation between the waist locations of the driver (focused upstream) and the trailing bunch (focused downstream) remained fixed at 21.3 mm, and that their waist beta functions stayed approximately constant throughout the scan.

The emittance was observed to increase dramatically when the beam was focused upstream of the plasma-cell entrance (see Fig. 4a). Conversely, when focused downstream, the emittance stayed approximately constant. However, in this case, significant loss of charge was observed for beam-waist locations beyond +10 mm (see Fig. 4c). This asymmetric behaviour may be explained by the accompanying change in driver focusing: focused upstream, the lower-density driver takes longer to establish a blowout cavity, which initially causes the trailing bunch to experience nonlinear focusing; focused downstream, the emittance-preserving blowout cavity is established immediately, but the large beam size of the mismatched trailing bunch causes it to lose charge from defocusing in the cavity walls.

Ultimately, locating the emittance-preserving operating point (see Fig. 2) required a fine-tuning of the quadrupoles used for alignment and matching at the level of 0.2% and 0.1%, respectively.

## Preservation of beam brightness

A more unified comparison of all beam qualities before and after acceleration can be made using the projected three-dimensional (3D) beam brightness, calculated by dividing the peak spectral density by the projected emittance (see "Methods"). The simultaneous preservation of emittance, charge, and energy spread implies that the 3D beam brightness in the horizontal plane was preserved (see Fig. 4e); 39% of shots in a 9-mm range of waist loca-tions fell within the $68^{\text{th}}$ percentile range of the initial brightness. Moreover, bunch lengthening does not typically occur within a plasma accelerator, implying that the 4D brightness was also likely to have been preserved. Lastly, although it could not be measured in this experiment, the axial symmetry of a plasma accelerator suggests that emittance preservation can also be demonstrated in the vertical plane (Methods), ultimately resulting in full 6D beam-brightness preservation.

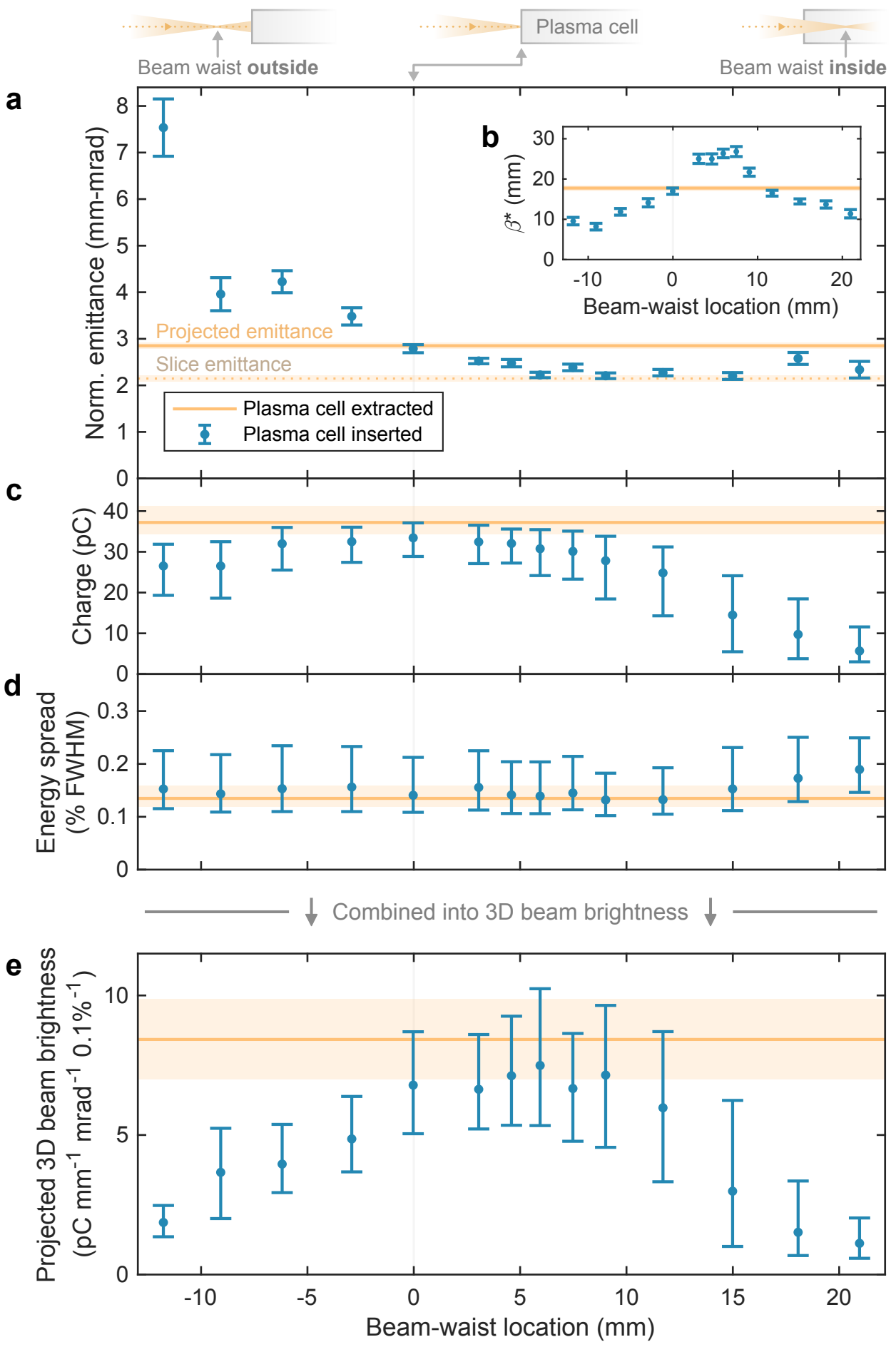

**Fig. 4 | Evolution of beam qualities and 3D brightness with mismatching. a**, The projected emittance (blue error bar) increased when the beam was focused upstream of the plasma-cell entrance. Focused downstream, the initial emittance (orange line) was preserved and even reduced down to the initial slice emittance (orange dotted line). **b**, Throughout the scan, the virtual-waist beta function $\beta^*$ varied significantly, consistent with a change in matching. **c**, With the beam waist at the plasma-cell entrance, the charge was preserved for roughly half the shots (see Fig. 2d), whereas as the waist moves away from the entrance, charge is progressively lost. **d**, The energy spread was similarly preserved for most scan steps. **e**, Combining the above beam qualities, the projected 3D beam brightness was preserved for beam-waist locations in the range from 0 to +9 mm. In **a–b**, error bars represent the best-fit value and fit error, whereas in **c–e** they represent the median and $68^{\text{th}}$ percentile range of the shot distributions.

## Discussion

While the emittance preservation achieved in this experiment (see Fig. 2) was associated with modest energy gain, the plasma accelerator was sufficiently long to require the same techniques and level of precision in the alignment and matching (see Figs. 3 and 4) that would also be required in longer plasma cells necessary for larger energy gain.

Whether emittance preservation can be achieved for larger energy gain can be tested via simulation. Starting from a PIC simulation that agrees with our experimental results (see Supplementary Fig. 10), we simulate identical input beams and plasma-density ramps but extend the central flat-top density to 500 mm. This results in significantly more energy gain (~700 MeV) while still preserving the emittance to within the measurement error (see “Methods” and Supplementary Fig. 11).

Reaching even larger energy gain can be accomplished either by further increasing the length of the stage or by using multiple stages[37]. For much longer acceleration distances, the main challenge will be to suppress transverse instabilities. The hose instability of the driver can be mitigated by increasing its transverse beam size[35], as in this experiment, and further suppressed by the large energy spread induced by deceleration[38]. However, the trailing bunch may still develop a beam-breakup instability[39,40], especially if operated at high energy-transfer efficiency. While beam breakup was not a problem for our parameters—the centroid offset can be estimated[41] to grow by less than ~0.5% (or ~86% for an extended stage with ~700 MeV energy gain)—it can in general be mitigated by detuning the betatron-oscillation frequency of different slices within the bunch. Methods include introducing an energy chirp[42] or a controlled amount of ion motion to provide nonuniform focusing within the bunch[43,44]—strategies that must be carefully balanced against their potential for additional emittance growth. Connecting multiple plasma-accelerator stages, on the other hand, presents another set of challenges for emittance preservation[45]; including chromaticity[46], distortion from in- and out-coupling devices[47], as well as nonlinear fields in compact focusing devices such as active plasma lenses[48–50].

In summary, we have demonstrated that beam quality, and in particular the emittance, can be preserved in a plasma-wakefield accelerator stage. This is a crucial step toward compact, high-energy particle accelerators for applications such as high-brightness FELs or high-luminosity linear colliders, where the performance critically depends on emittance and other beam qualities.

## Methods

### Electron driver- and trailing-bunch generation

The FLASH linac provided electron bunches with 880 pC of charge from a photocathode source, accelerated to 1050 MeV by superconducting RF cavities. The bunches were compressed with two magnetic chicanes to a bunch length of 285 μm rms, and approximately linearized in longitudinal phase space with a third-harmonic cavity. Active feedbacks for charge, energy, bunch length, and orbit were used to stabilize the operation over the multi-hour data-acquisition period. A double-bunch temporal structure was created by dispersing the electrons in energy in the horizontal plane onto two block collimators[33] that removed the high- and low-energy tails, as well as a notch collimator that split the bunch into a leading driver (400 pC) and a trailing bunch (40 pC).

### Transverse alignment and final focusing

Horizontal alignment of the two bunches was accomplished by adjusting a quadrupole and a sextupole, located downstream of the collimators in a region of large horizontal dispersion, in order to cancel first- and second-order tilts[51]. Vertical alignment was not critical, as negligible vertical dispersion was introduced through-out the beam line. Final focusing was performed using nine quadrupoles in a 13 m-long straight section just downstream of the dispersive section. These quadrupoles were optimized to focus the beam to small beta functions while minimizing the first-order chromaticity in both planes (Supplementary Fig. 1). The longitudinal position of the beam waist, located close to the plasma-cell entrance, was precisely adjusted in the horizontal plane using the third-last quadrupole before the plasma cell, where the horizontal beta function was 6.8 times larger than in the vertical plane, and adjusted in the vertical plane using the second-last quadrupole, where the vertical beta function was 7 times larger than in the horizontal plane. While the beam was strongly focused and the beam current was high, space-charge effects were nevertheless negligible due to the GeV-level particle energy.

**Beam-waist measurements**

The location and beta function of the beam waist, as well as the relative misalignment between the driver and the trailing bunch, were estimated using a BPM-based measurement technique developed in a previous publication[36], where the multi-shot distribution of the orbit jitter is used as a proxy for the beam. Supplementary Figure 2 shows these beam-waist parameters measured at each step of the scans shown in Fig. 3 (horizontal alignment scan) and Fig. 4 (horizontal beam-waist location scan). The fit in Supplementary Fig. 2c was used for angular calibration in Fig. 3, and the fit in Supplementary Fig. 2d was used for the calibration of beam-waist locations in Fig. 4.

**Longitudinal-phase-space measurements**

The charge distribution of the electron bunches in longitudinal phase space was characterized using a PolariX-type[52] X-band RF transverse-deflection structure (TDS) placed 33 m downstream of the plasma cell. During this measurement, no beam–plasma interaction took place. The electron bunches were streaked vertically by the TDS and horizontally dispersed by a dipole magnet onto an in-vacuum GAGG:Ce screen. In order to maximize the resolution, three quadrupole magnets were used to point-to-point image in the dispersive plane, from the TDS to the measurement screen, and parallel-to-point image in the streaking plane. Supplementary Figure 3 shows the longitudinal phase space of the double-bunch structure using a two-point tomographic reconstruction[53] based on both zero crossings in order to remove distortions caused by dispersion. This reconstruction shows that the driver had an average peak current of 1.0 kA and a bunch length of 42 μm rms (140 fs rms), while the trailing bunch had an average peak current of 0.44 kA and a bunch length of 11 μm rms (37 fs rms). The centroids of the two bunches were separated by 195 μm (650 fs). Individual shots (Supplementary Fig. 3a–b) show evidence of a microbunching instability[54]; however, these microbunches do not significantly affect the plasma wake as their wavelength of ~10 μm (~33 fs) is much shorter than the minimum plasma wavelength of ~300 μm (~1 ps).

**Plasma generation and density measurements**

The plasma was generated using a discharge capillary[55], consisting of a 1.5 mm-diameter, 50 mm-long channel milled from two sapphire blocks. Argon gas doped with 3% hydrogen continuously flowed into the capillary via two gas inlets, located 2.5 mm from the entrance and exit. Using a backing pressure of 30 mbar, the resulting capillary pressure was 9 mbar, as measured with a pressure sensor connected close to the gas inlet. Short (400 ns flat-top), high-voltage (25 kV), high-current (400 A) discharge pulses between two electrodes at the capillary entrance and exit were used to ignite the plasma, after which the density decayed exponentially with a half-life of 2.1 μs (Supplementary Fig. 4a). This was measured using spectral-line broadening of the H-alpha line[56], observed with an optical spectrometer collecting light from the full capillary radius in a 7-mm longitudinal region near the centre of the plasma cell, integrating over 0.2 μs on an intensified camera. The beam arrived 9.68 μs after the initial discharge, at which time the radially averaged plasma density at the centre of the plasma cell had decayed to approximately $8\times10^{15}$ cm$^{-3}$. The longitudinal plasma-density profile, measured by displacing the cell longitudinally[57], is consistent with a Gaussian-like density profile (Supplementary Fig. 4b). The radial plasma-density profile was not measured, but the on-axis density during the period of exponential decay (beyond ~2 μs) is expected to be approximately 50% higher than the measured average (i.e., $1.2\times10^{16}$ cm$^{-3}$); this effect is observed when electron beams are translated radially inside the capillary, as well as in magnetohydrodynamic simulations[58]. Low-density longitudinal ramps outside the plasma cell, which can affect the beta function, could also not be measured. Nevertheless, as a best estimate for PIC simulations, we assumed an inverse-square profile (reaching half density 4 mm outside the electrodes of the cell) based on observed cone-shaped light emissions.

**Broad-band and high-resolution imaging spectrometers**

Two electron-energy spectrometers were used in this experiment, one for broad-band spectrum measurements and another for the high-resolution emittance measurements, both using a 1.07 m-long vertically dispersive dipole magnet. Five quadrupole magnets with a 5 mm bore radius were used to point-to-point image the diverging electron bunches from the plasma cell to the measurement screens. The broad-band spectrometer used an out-of-vacuum scintillator screen (GadOx), giving a spatial resolution of approximately 50 μm rms, placed 4 m downstream of the plasma cell, resulting in a horizontal beam-imaging magnification of a factor -3. The high-resolution spectrometer used an in-vacuum scintillator screen (GAGG:Ce) located 7.3 m downstream of the plasma cell, resulting in a larger horizontal magnification of -7.9. The corresponding vertical magnification was approximately -2.6 (for ~1.05 GeV) and -2.7 (for ~1.1 GeV). The resolution of this screen (part of an European XFEL-type screen station[59]), imaged with Scheimpflug optics, was measured to be 6.2 μm rms or smaller (Supplementary Fig. 5) by imaging a beam focused to less than 5 μm rms using an imaging optic with a magnification of -1. The pixel size, corresponding to 5.5 × 5.5 μm$^2$ on the screen, does not contribute significantly to the resolution, but was nevertheless accounted for as part of the above resolution measurement. A charge-density calibration was performed on both screens by scanning the position of an energy collimator and correlating the integrated on-screen charge with that measured in an upstream toroidal current transformer. Scintillator saturation effects in the high-resolution screen were accounted for by correcting the light yield by Birk's law[60]

$$\rho = \frac{\rho_{\mathrm{scint}}}{1 - B\rho_{\mathrm{scint}}},$$

where $\rho$ is the true charge density, $\rho_{\mathrm{scint}}$ is the apparent charge density measured on the scintillator and $B$ is Birk's constant. The material GAGG:Ce was chosen specifically for its high saturation threshold[61], which was estimated experimentally to be approximately 18 nC mm$^{-2}$, or equivalently to a $B$ = 0.056 mm$^2$ nC$^{-1}$. However, since the peak charge density obtained during emittance measurements (i.e., while point-to-point imaging the virtual waist) was never above 2.7 nC mm$^{-2}$, this correction is small and has only a percent-level effect on the average measured charge.

**Energy-transfer-efficiency measurements**

The energy-transfer efficiency was calculated to be 15–25% (with a distribution mode of 22%) by comparing the energy gained by the trailing bunch to the energy lost by the driver,

$$\eta = -\frac{\Delta E_{\text{trailing}} Q_{\text{trailing}}}{\Delta E_{\text{driver}} \bar{Q}_{\text{driver}}},$$

where $\Delta E$ denotes the change in mean energy, and $Q$ is the charge: the final charge of the trailing bunch, and the mean of the measured initial and final charge for the driver (i.e., a best estimate in case of missing driver charge after deceleration). The distribution of efficiency is shown in Supplementary Fig. 6. For improved accuracy, the average driver spectrum was reconstructed from a scan of imaging energies, making use only of the part of the spectrum closest to the correctly imaged energy.

**Emittance measurements**

The emittance of the trailing bunch was measured by scanning the strength of the imaging quadrupoles such that only the horizontal object plane changed, whereas the horizontal magnification and vertical object plane (for good energy resolution) remained constant, as shown in Supplementary Fig. 7. A slight vertical offset of one or more of these quadrupoles led to a slight shift in the apparent energy throughout these object-plane scans (as seen in the energy projection in Fig. 2b). The object plane and magnification were calculated for each shot individually based on the measured mean energy of the trailing bunch and the current in each quadrupole. The true beam size (i.e., at the location of the object plane) was calculated by dividing the measured beam size on the screen by the magnification. The screen resolution had a negligible effect, increasing the measured beam size by 1.5% or less. The normalized emittance $\varepsilon_n$, waist beta function $\beta^*$, and beam-waist location $s^*$ were extracted by fitting a ballistic beam-waist envelope model to the measured true beam size

$$\sigma(s) = \sqrt{\frac{\epsilon_n}{\gamma}\left(\beta^* + \frac{(s-s^*)^2}{\beta^*}\right) + \left(\frac{\sigma_{\text{res}}}{M}\right)^2},$$

where $s$ denotes the object plane, $\sigma_{\text{res}}$ denotes the screen resolution and $M$ denotes the beam-imaging magnification. The exact magnification was $M$ = -7.87 (with 0.03% rms jitter) with the plasma cell extracted (see Fig. 2b), and $M$ = -7.88 (with 0.15% rms jitter) with the plasma cell inserted (see Fig. 2c), where the increased jitter is caused by the energy jitter resulting from the plasma acceleration. For the emittance measurements used for Figs. 3 and 4 (shown in full in Extended Data Figs. 8 and 9), the incoming bunch length and the beam charge (combined driver and trailing bunch charge) was filtered to only include a range ±15% and ±5%, respectively, to ensure similar input parameters throughout the multi-hour measurement. Here, the bunch length was measured prior to double-bunch generation (i.e., notch collimation) with a calibrated pyroelectric detector for coherent diffraction radiation. Lastly, in addition to the measurement uncertainty, the incoming projected emittance shown in Fig. 3 has an uncertainty related to dispersion induced by misalignment: $D_{x'} \approx \Delta x'/(\delta E/E)$, where $\Delta x'$ and $\delta E/E$ = 0.9% are the relative angle and relative energy difference between the driver and trailing bunches, respectively. This dispersion, when multiplied by the relative energy spread $\sigma_\delta \approx 0.06\%$ rms of the trailing bunch, adds/subtracts in quadrature with the measured divergence $\sigma_{x'}$ = 0.28 mrad rms, resulting in a relative emittance uncertainty

$$\frac{\sigma_\epsilon}{\epsilon} = \sqrt{1 + \left(\frac{\sigma_\delta D_{x'}}{\sigma_{x'}}\right)^2} - 1,$$

corresponding to a 6% added uncertainty at maximal misalignment ($\Delta x'$ = ±1.2 mrad).

**Projected 3D-beam-brightness calculations**

The projected 3D beam brightness, as shown in Fig. 4e, is calculated using the formula

$$B_{3\text{D}} \equiv \frac{1}{\epsilon_{nx}}\left(\frac{\partial Q}{\partial \delta}\right)_{\text{peak}},$$

where the peak of the relative spectral charge density, $\partial Q/\partial\delta$, is divided by the projected normalized emittance in the horizontal plane, $\varepsilon_{nx}$. Here, $\delta = \Delta E/E$ is defined as the relative energy offset. The uncertainty of the 3D brightness, shown as error bars in Fig. 4e, is estimated by Monte-Carlo sampling: dividing the peak spectral density of all the shots in each step by a large number of normally distributed samples of the emittance in that step (whose distribution is defined by the best fit value and error), and then quantifying the width of the resulting 3D brightness distribution as the 68$^{\text{th}}$ percentile range (equivalent to ±1 sigma if the distribution would be normal, which it is not).

**Particle-in-cell simulations**

Particle-in-cell simulations were performed using the open-source 3D code HiPACE++[62], which uses the quasi-static approximation. The input beam was reconstructed in 6D phase space based on beam-waist measurements using BPMs (Supplementary Fig. 2), longitudinal-phase-space measurement using a TDS (Supplementary Fig. 3), as well as the measured transverse phase space of the trailing bunch (see Fig. 2). The horizontal and vertical slice emittances of the driver were not measured, but kept as free parameters. The longitudinal plasma-density profile was based on the optical spectrometer measurement, with assumed external ramps (Supplementary Fig. 4). Since the incoming vertical emittance of the trailing bunch could not be measured on the spectrometer, it was assumed to be identical to that of the incoming horizontal emittance—roughly consistent with previous measurements elsewhere in the linac. Simulations were performed in a box of size 1024 × 1024 × 600 μm$^3$ in the horizontal, vertical, and longitudinal directions, respectively, with 0.5 μm resolution (i.e., 2048 × 2048 × 1200 grid cells). The step size was 195 μm (650 fs). The beam was resolved with 4 million constant-weight macroparticles; the plasma was resolved with 4 particles per cell. The simulation results (Supplementary Fig. 10) are consistent with all the experimental measurements: the charge and the projected normalized emittance in the horizontal plane are both preserved, while the energy spread was slightly reduced. The simulation also suggest that the emittance was (or could in principle be) preserved also in the vertical plane.

**Particle-in-cell simulation with larger energy gain**

In order to assess the scalability of the measured emittance preservation, a PIC simulation with an artificially extended plasma cell was performed. Here, both the input beam parameters, as well as the plasma-density profile in the up and down ramps, are identical to that used in the shorter simulation (shown in Supplementary Fig. 10) in order to keep the same matching as in the experiment. However, a central 500-mm flat-top density region has been introduced (between the up and down ramps) to increase the energy gain to approximately 700 MeV per particle in the trailing bunch. To speed up the simulation, the simulation box size was reduced (to $512 \times 512 \times 400$ μm$^3$), while keeping the same resolution, step size and number of beam particles. A small linear beam tilt of 0.4% (i.e., a 1 μm transverse offset per 250 μm longitudinal offset behind the head of the driver) is introduced in both the horizontal and vertical planes—small enough to not affect the initial beam quality or dynamics, but sufficient to seed exponentially growing transverse instabilities. The results are shown in Supplementary Fig. 11. Here, the emittance in the horizontal is observed to increase by only 0.5%—preserved to well within the measurement error (±3%). In the vertical plane, the simulation indicates an emittance growth of 2.5%—somewhat higher, as the matching was less optimized. While the trailing bunch undergoes 8020° of phase advance (i.e., nearly 45 betatron envelope oscillations), no transverse instabilities are observed to cause emittance growth. Coulomb scattering was not included in this simulation, but can be estimated analytically to increase the emittance by approximately $1.2\times10^{-3}$ mm-mrad, which is negligible. We observe that in this simplistic extension, the peak spectral density is not preserved, as the current profile of the trailing bunch is not optimized for the flat-top region that dominates the longer cell, but was rather a result of experimental optimization with respect to the longitudinally averaged wakefield in the shorter cell. Consequently, the trailing bunch over-loads the wakefield in extended flat-top region, leading to higher energy-transfer efficiency (40%) but also a chirped distribution in longitudinal phase space with a 2.6% rms energy spread. This can be avoided in future experimental demonstrations with larger energy gain by re-optimizing the trailing-bunch current profile and/or the properties of the driver (to drive a stronger wakefield), such that the wakefield is uniform and the energy spread and peak spectral density remain preserved.

## Data availability

All data are available upon reasonable request from the corresponding authors.

## Code availability

HiPACE++ is openly available (https://github.com/Hi-PACE/hipace) and all other codes and tools are available upon reasonable request.

# Acknowledgements

We thank M. Dinter, S. Karstensen, S. Kottler, K. Ludwig, F. Marutzky, A. Rahali, V. Rybnikov, A. Schleiermacher, the FLASH management, and the DESY FH, M and FS divisions for their scientific, engineering and technical support. This work was supported by Helmholtz ARD, Helmholtz ATHENA, and the Helmholtz IuVF ZT-0009 programme, the Maxwell computational resources at DESY, as well as the Research Council of Norway (NFR Grant No. 313770). The authors gratefully acknowledge the Gauss Centre for Supercomputing e.V. (www.gauss-centre.eu) for funding this project by providing computing time through the John von Neumann Institute for Computing (NIC) on the GCS Supercomputer JUWELS at Jülich Supercomputing Centre (JSC).

# Author contributions

C.A.L., R.D., and J.O. conceived the experiment. C.A.L., L.B., J.B.S., F.P. and J.C.W. performed the experiment, with help from J.B., J.C., R.D., J.M.G., P.G.C., G.L., S.S., and S.W. J.M.G and G.L. performed the plasma-density measurements. C.A.L. analyzed the experimental data and produced all the figures. C.A.L. wrote the manuscript, with assistance from B.F. C.A.L. performed the 6D beam reconstruction and corresponding HiPACE++ PIC simulations, with help from S.D. and M.T., who also developed new HiPACE++ features for these simulations. R.D. and J.O. supervised the project and the personnel. J.B. and J.C. were supervised by M.W. All authors discussed the results in the paper.

# Competing interests

The authors declare no competing interests.

# Additional information

**Supplementary information** The online version contains supplementary material.

**Correspondence** and requests for materials should be addressed to C. A Lindstrøm, R. D'Arcy or J. Osterhoff.

# Supplementary Figures

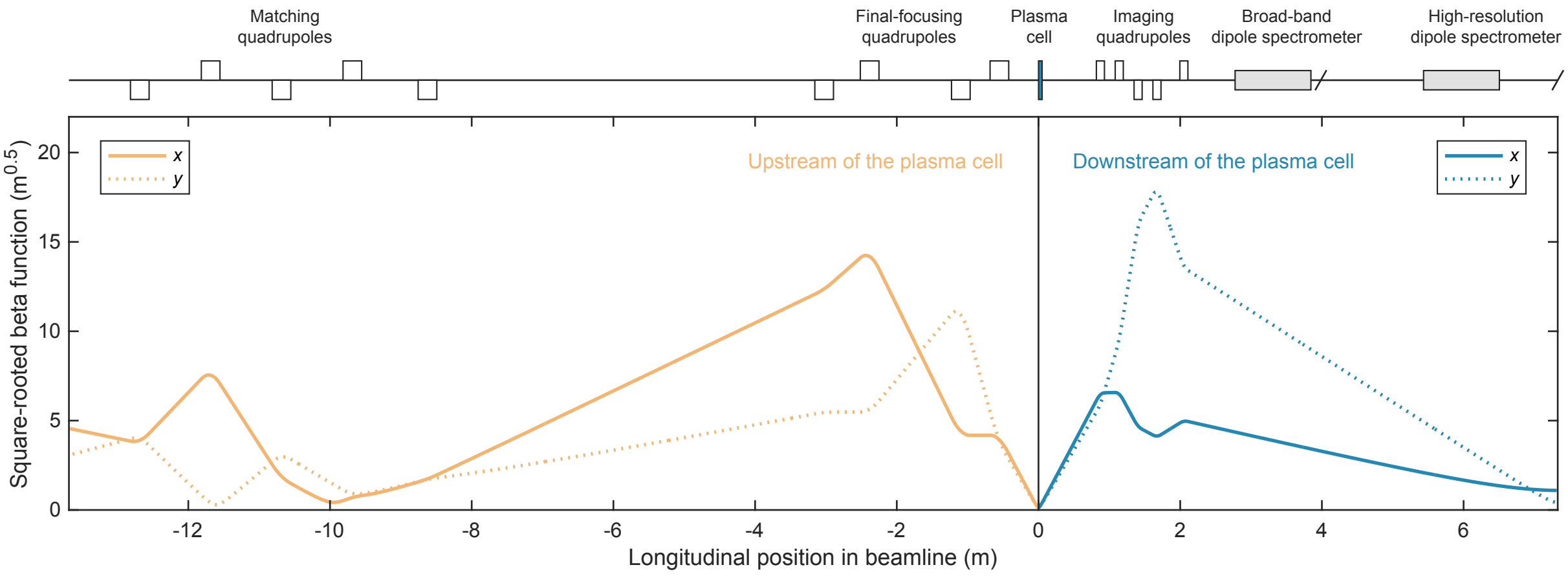

**Supplementary Fig. 1 | Final-focusing and imaging optics.** The electron bunches were strongly focused to match the beam size and beta function to the plasma-wakefield accelerator. Nine quadrupole magnets (white rectangles in the above beam-line sketch) were used for matching and final focusing upstream of the plasma cell (dark blue rectangle in the sketch). Here, the square root of the beta function in the horizontal (solid lines) and vertical (dotted lines) planes are shown; this quantity is proportional to the beam size. Downstream of the plasma cell, five quadrupoles were used for capturing the diverging electron bunches and point-to-point imaging them onto one of two spectrometer screens (black slanted lines in the sketch) after dispersion by a dipole magnet (grey rectangles in the sketch). Only the optics for the high-resolution-spectrometer screen, used for emittance measurements, is shown.

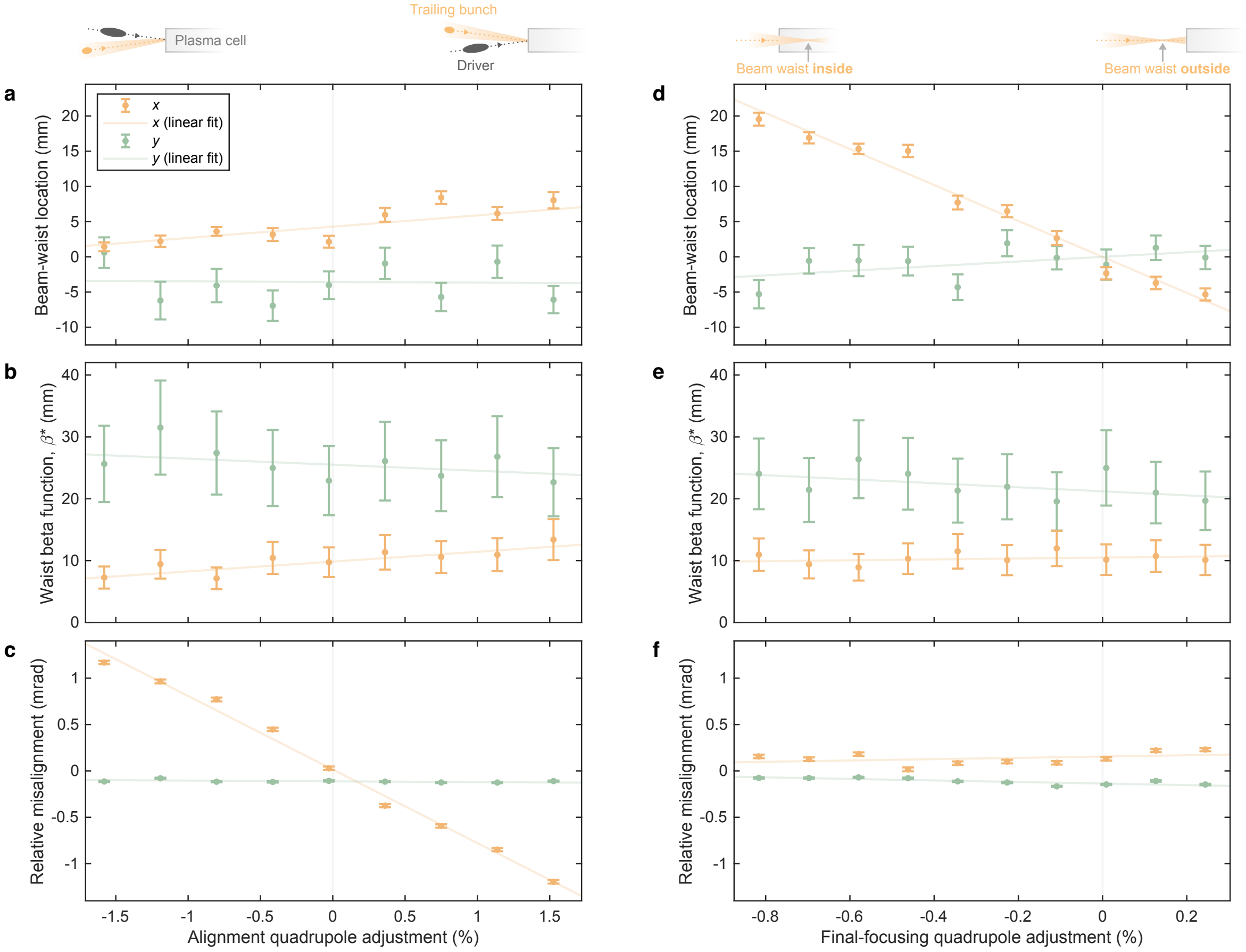

**Supplementary Fig. 2 | Relative changes of the beam waist and alignment estimated using beam-position monitors.** The distribution of beam trajectories jittering over multiple shots were used to measure relative changes in beam-waist location and beta function[36], as well as the misalignment between the driver and trailing bunches. **a**, The beam-waist location was measured for each step in the alignment scan used in Fig. 3, where the beam-waist location of the emittance-preserving operating point (Fig. 2) is defined as zero. A small relative shift in the horizontal plane (orange error bars) occurred when adjusting the strength of the alignment quadrupole (moves 1.6 mm downstream per 1%), while remaining approximately constant in the vertical plane (green error bars). In all plots, the points and error bars represent the best fit value and statistical uncertainty; 150 shots were used per data point. **b**, A similar change occurred for the horizontal-waist beta function (1.6 mm larger per 1%), while the vertical-waist beta function remained approximately constant. **c**, The relative alignment between the driver and trailing bunches, whose trajectories were measured separately, changed significantly in the horizontal plane (0.8 mrad per 1%), while remaining unchanged in the vertical plane. **d–f**, A similar measurement was performed for the scan of the beam-waist location shown in Fig. 4. Here, the horizontal-waist location changed significantly when changing the strength of a final-focusing quadrupole (moving 25 mm upstream per 1%), while the vertical-waist location moved negligibly (moving 3 mm downstream per 1%). The waist beta functions and relative misalignments also remained approximately constant.

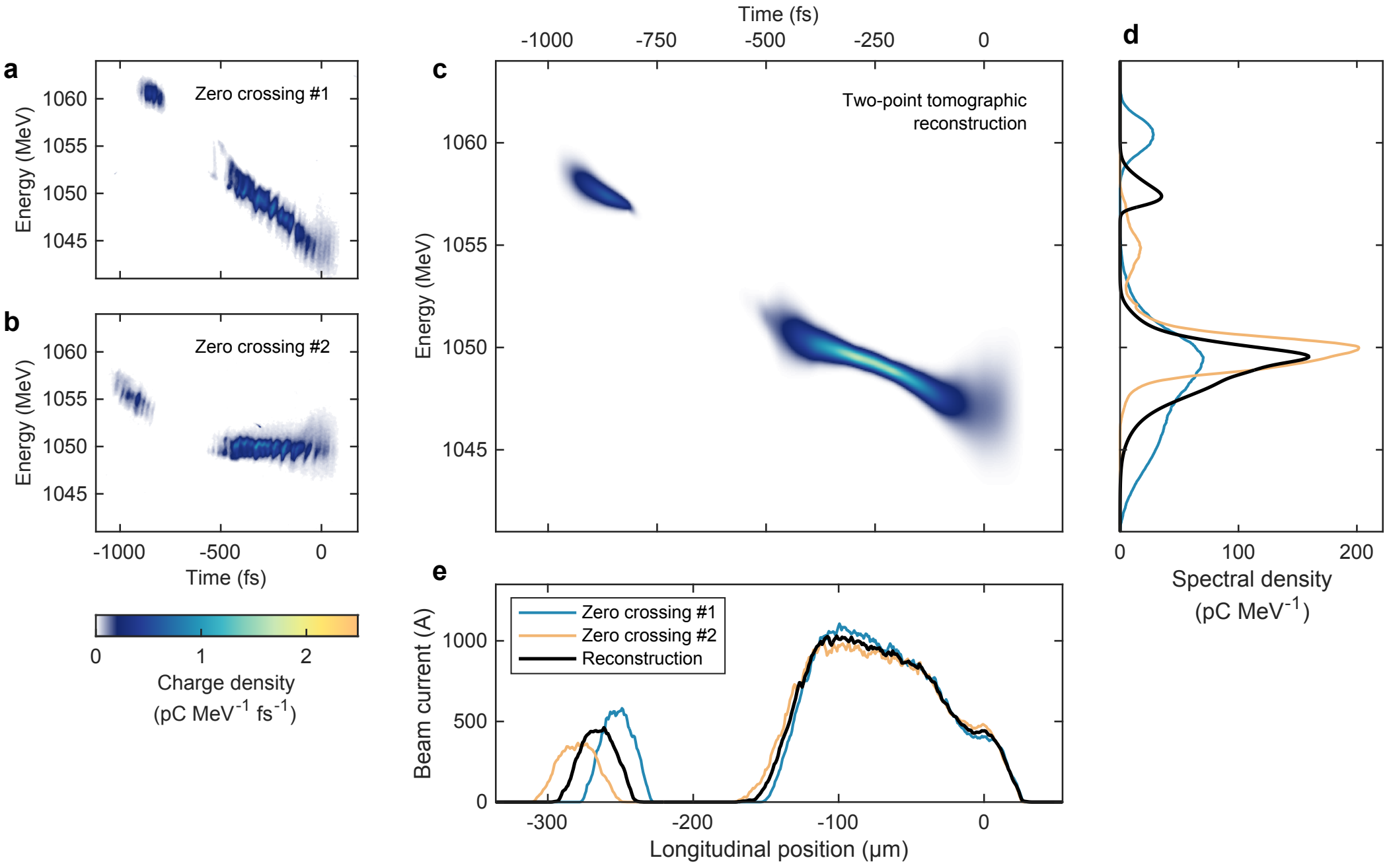

**Supplementary Fig. 3 | Longitudinal-phase-space measurement.** A transverse-deflection structure and a magnetic dipole were used to streak in time and disperse in energy, respectively, onto a scintillator screen. **a–b**, The measurement was performed in both zero crossings of the oscillating field. Microbunches can be observed in the streaked bunches. **c**, Using 50 shots from each zero crossing, a two-point tomographic reconstruction can be made; no microbunching is visible due to multi-shot averaging. Two bunches, a driver (bottom right) and a trailing bunch (top left), were created using a notch collimator in an energetically dispersive region. Here, the energy resolution of 0.64 MeV rms was removed by subtraction in quadrature. **d**, An energy projection shows that the two bunches are spectrally distinct. **e**, Similarly, the measured current profile indicates that the driver and trailing bunches are temporally distinct, with peak currents of 1.0 and 0.44 kA, respectively.

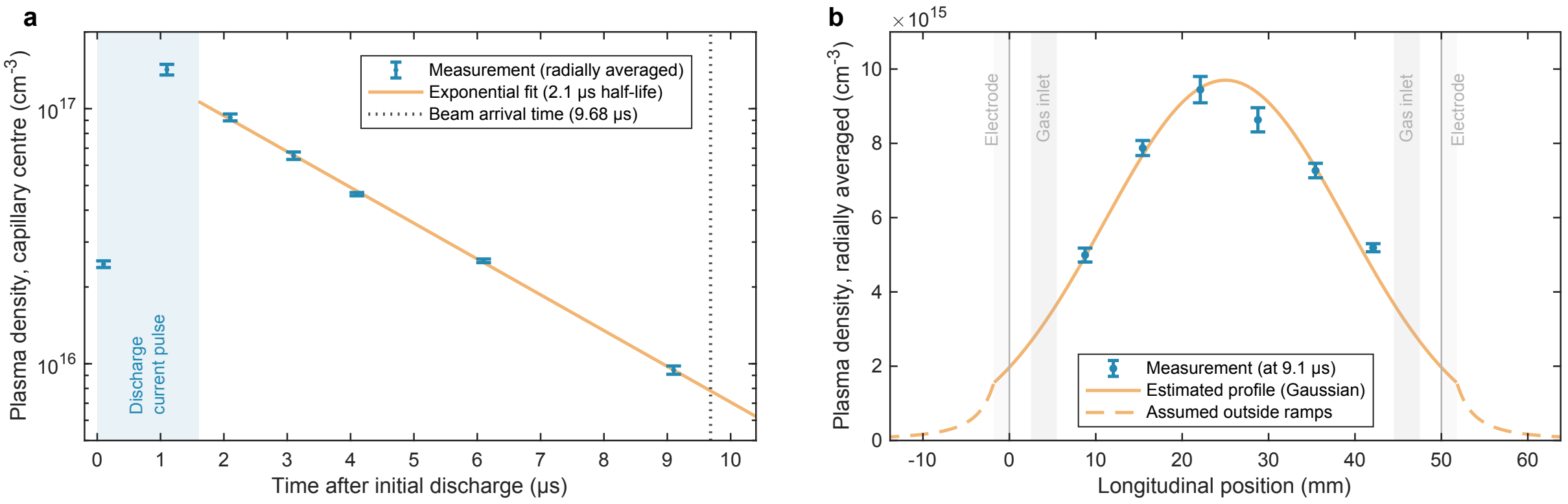

**Supplementary Fig. 4 | Plasma-density evolution and longitudinal profile. a**, The plasma density at the capillary centre (blue error bar) was measured based on H-alpha spectral-line broadening, where the error bars indicate the statistical fit error. This was integrated across a 7 mm-long region longitudinally and across the full radius of the capillary. After the discharge current pulse (blue area) stops, the central plasma density decays exponentially with a half-life of 2.1 μs (orange line). The beam arrived 9.68 μs after the initial discharge. **b**, Close to this time (at 9.1 μs) the measured longitudinal plasma-density profile (blue error bars) has a Gaussian-like shape inside the cell (orange curve). It was not possible to measure the low-density ramps outside the electrodes of the plasma cell—a polynomial shape was assumed in PIC simulations, emulating a cone-shaped expulsion. At the beam arrival time (0.58 μs later), the density profile is expected to retain its shape, but decay by a further 20% (to a peak of $8\times10^{15}$ cm$^{-3}$). The on-axis density, as experienced by the beam, is expected to be approximately 50% higher than the measured radially averaged density (i.e., peaking at $1.2\times10^{16}$ cm$^{-3}$).

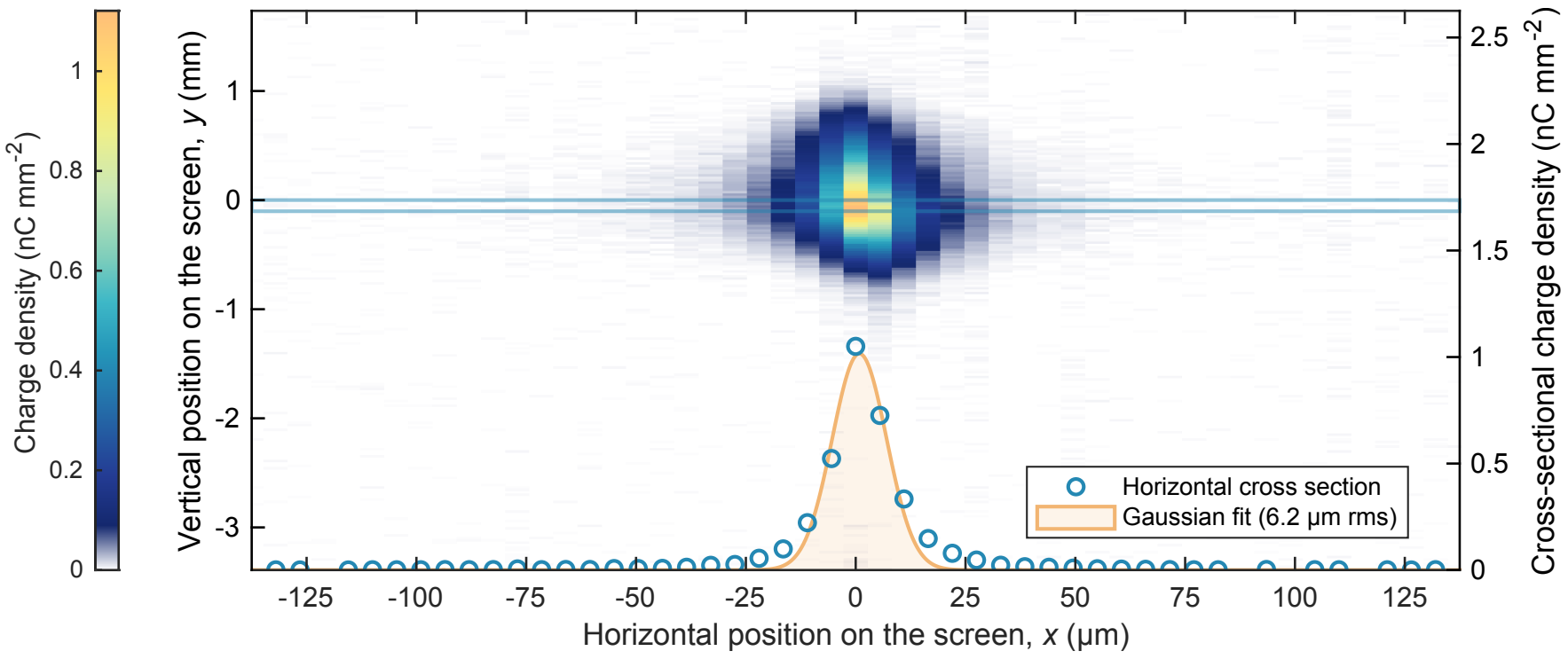

**Supplementary Fig. 5 | Screen-resolution measurement for the high-resolution spectrometer.** A tightly focused and highly collimated beam (with 20 pC of charge) was imaged onto the high-resolution spectrometer screen with a beam-imaging magnification of -1. Note that the image has unequal scales in the horizontal (bottom axis) and vertical planes (left axis). The central cross section in the horizontal plane (blue circles, right axis) shows that the point-spread function is Gaussian-like with extended tails. The pixel width corresponds to 5.5 µm on the screen. The fitted Gaussian distribution (orange area) has an rms of 6.2 µm, suggesting that the screen resolution is negligible compared to the beam sizes measured in the emittance measurements, which were operated with a magnification of -7.9.

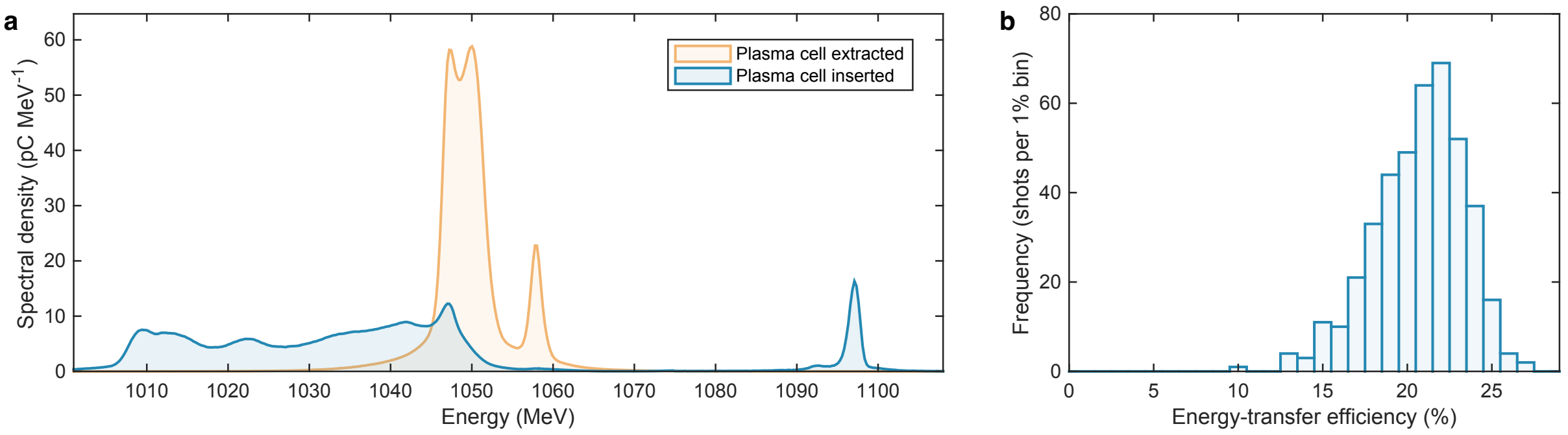

**Supplementary Fig. 6 | Measurement of energy-transfer efficiency. a**, Complete energy spectra were measured using the broad-band spectrometer, both while the plasma cell was extracted (orange area) and inserted (blue area). Here, the decelerated driver spectrum was reconstructed using an imaging-energy scan for increased accuracy. The accelerated trailing-bunch spectrum represents the shot with the highest peak spectral density. The energy spread appears higher (and spectral density lower) compared to Fig. 1f and Supplementary Fig. 3 due to a lower energy resolution, but this has a negligible effect on the calculated efficiency. **b**, Combining the energy lost by the driver (on average), as measured on the broad-band spectrometer, and the energy gained by the trailing bunch (shot-by-shot, not simultaneously), as measured on the high-resolution spectrometer (Fig. 2c–e), the energy-transfer efficiency of the emittance-preserving operating point can be calculated. A histogram shows the shot distribution of energy-transfer efficiency peaking at 22 ± 2.2%, and reaching a maximum of 27 ± 2.7%, where the quoted error arises from a systematic uncertainty of the driver energy loss, due to a 20% decrease in charge between the reconstructed cell-extracted and cell-inserted driver spectra.

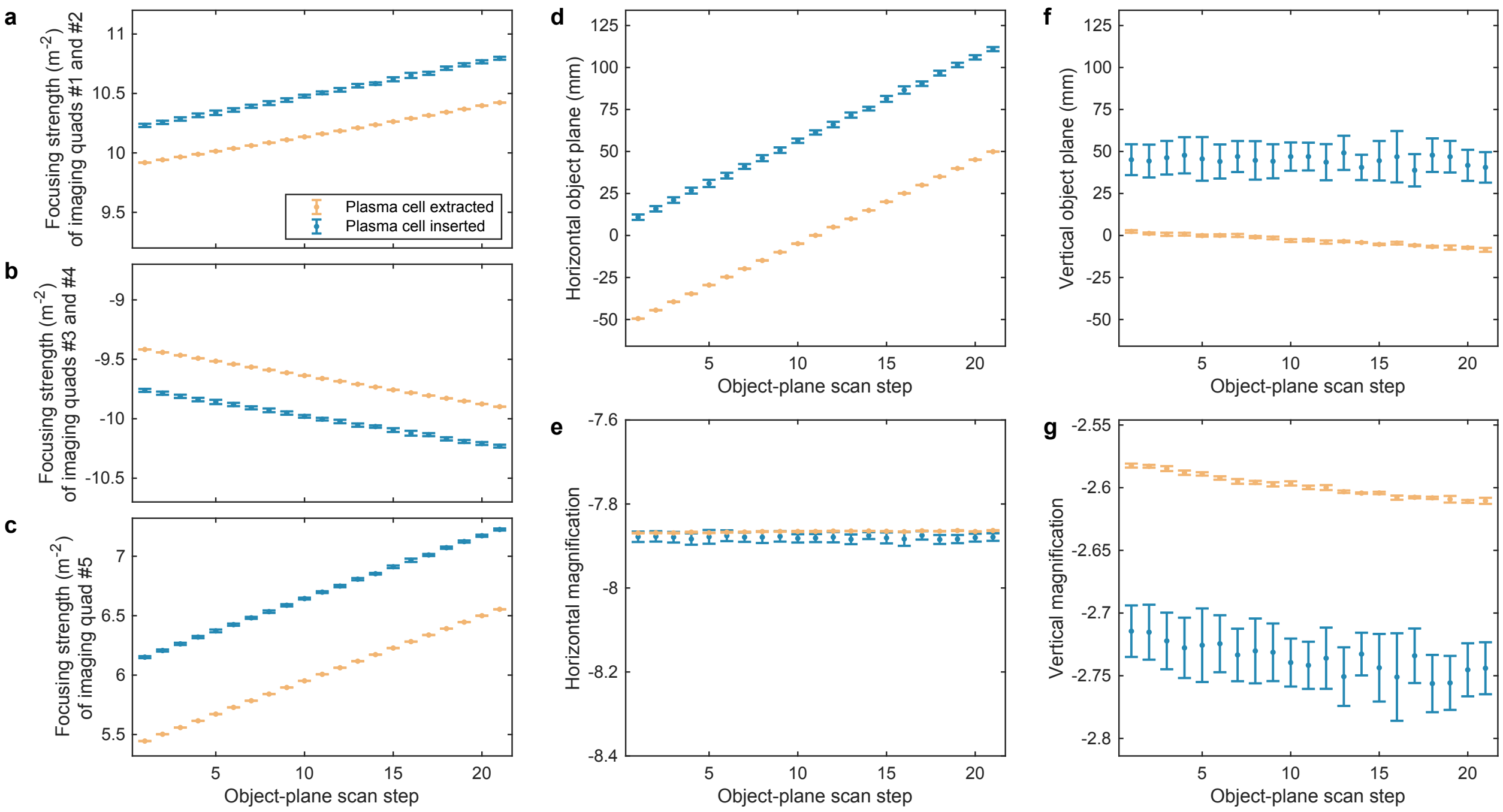

**Supplementary Fig. 7 | Object-plane scan for emittance measurements. a–c**, In order to scan the horizontal object plane while maintaining a constant horizontal magnification and constant vertical object plane, the current in each of the five imaging quadrupoles (three of which were independent) was calculated and precisely adjusted. The resulting focusing strengths, calculated for the mean trailing-bunch energy of each shot, are shown for the scans with plasma cell both extracted (orange error bars) and inserted (blue error bars), corresponding to Fig. 2b and 2c, respectively. In all plots, error bars indicate the mean and rms of the shots in each scan step. **d**, The object plane in the horizontal plane was scanned over a 100-mm range centred around the plasma entrance and exit. **e**, The horizontal magnification was kept constant at approximately -7.9. **f**, The vertical object plane was also kept constant at a location resulting in a small vertical beam size in the dispersive plane, to maintain high energy resolution. A small drift appears in the cell-extracted scan, but is sufficiently small to be negligible in the energy-spectrum measurement. **g**, Since only three quadrupole currents could be independently adjusted to satisfy the constraints on the horizontal and vertical object plane as well as the horizontal magnification, the vertical magnification could not be simultaneously controlled, but varied slightly around a value of -2.60 (plasma cell extracted) and -2.74 (plasma cell inserted).

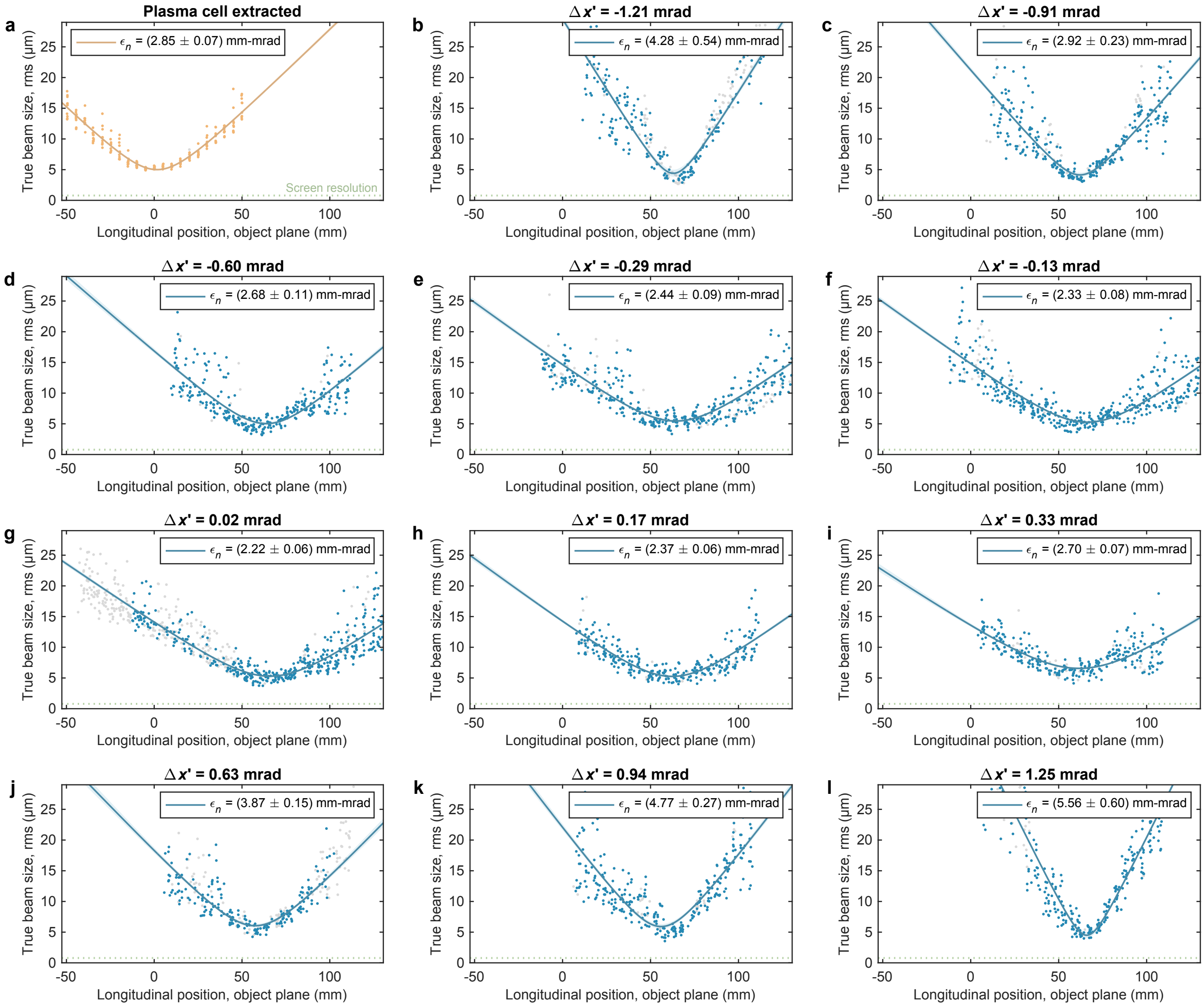

**Supplementary Fig. 8 | Emittance measurements for various misalignments. a**, An object-plane scan was performed while the plasma cell was extracted, measuring the beam size in the object plane for each shot (orange points). The quoted normalized emittance, $\varepsilon_n$, is found by a fit of the virtual waist (orange line). The screen resolution (green dotted line) is negligible. This is the same scan as shown in Fig. 2a. **b–l**, The measurement was repeated with the plasma cell inserted (blue points) for 11 separate values of $\Delta x'$, the angular misalignment between the driver and the trailing bunch. Shots are filtered (gray points) to ensure consistent input beams despite drift and jitter in bunch length (to be within ±15%) and charge (to be within ±5%), as well as a consistent scan range (to be within ±7.5 mm). These measurements correspond to the values in Fig. 3d.

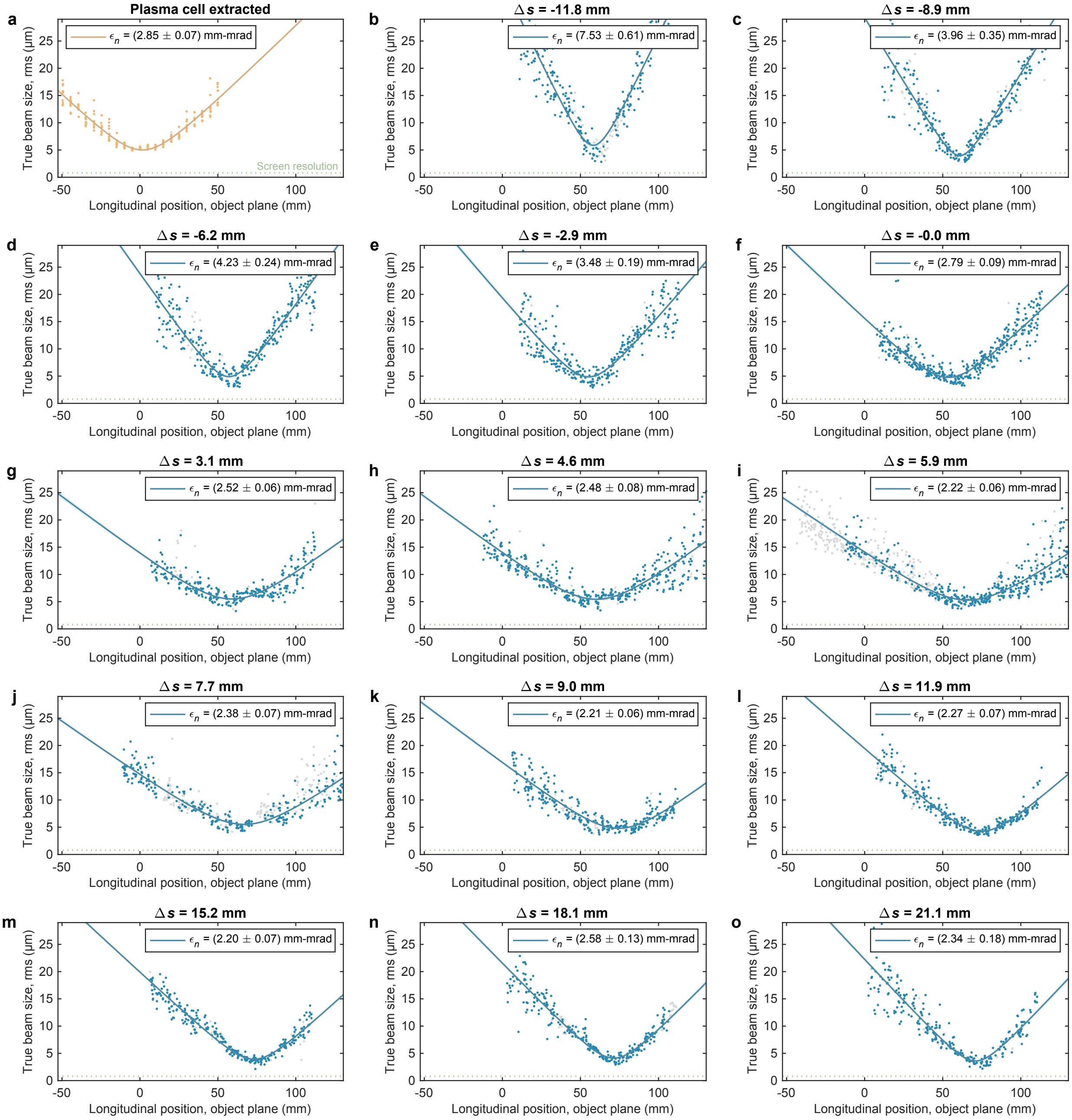

**Supplementary Fig. 9 | Emittance measurements for various beam-waist locations. a,** Similar to Supplementary Fig. 8, an object-plane scan was performed while the plasma cell was extracted, showing the measured beam size (orange points), the virtual-waist fit (orange line) and the negligible screen resolution (green dotted line). **b–o**, The beam size (blue points) was also measured with the plasma cell inserted for 14 different values of the trailing bunch's waist location $\Delta s$ relative to the plasma-cell entrance. Again, as in Supplementary Fig. 8, shots are filtered (gray points) to ensure consistency in bunch length (< ±15%), charge (< ±5%), and scan range (< ±7.5 mm). These measurements correspond to Fig. 4a.

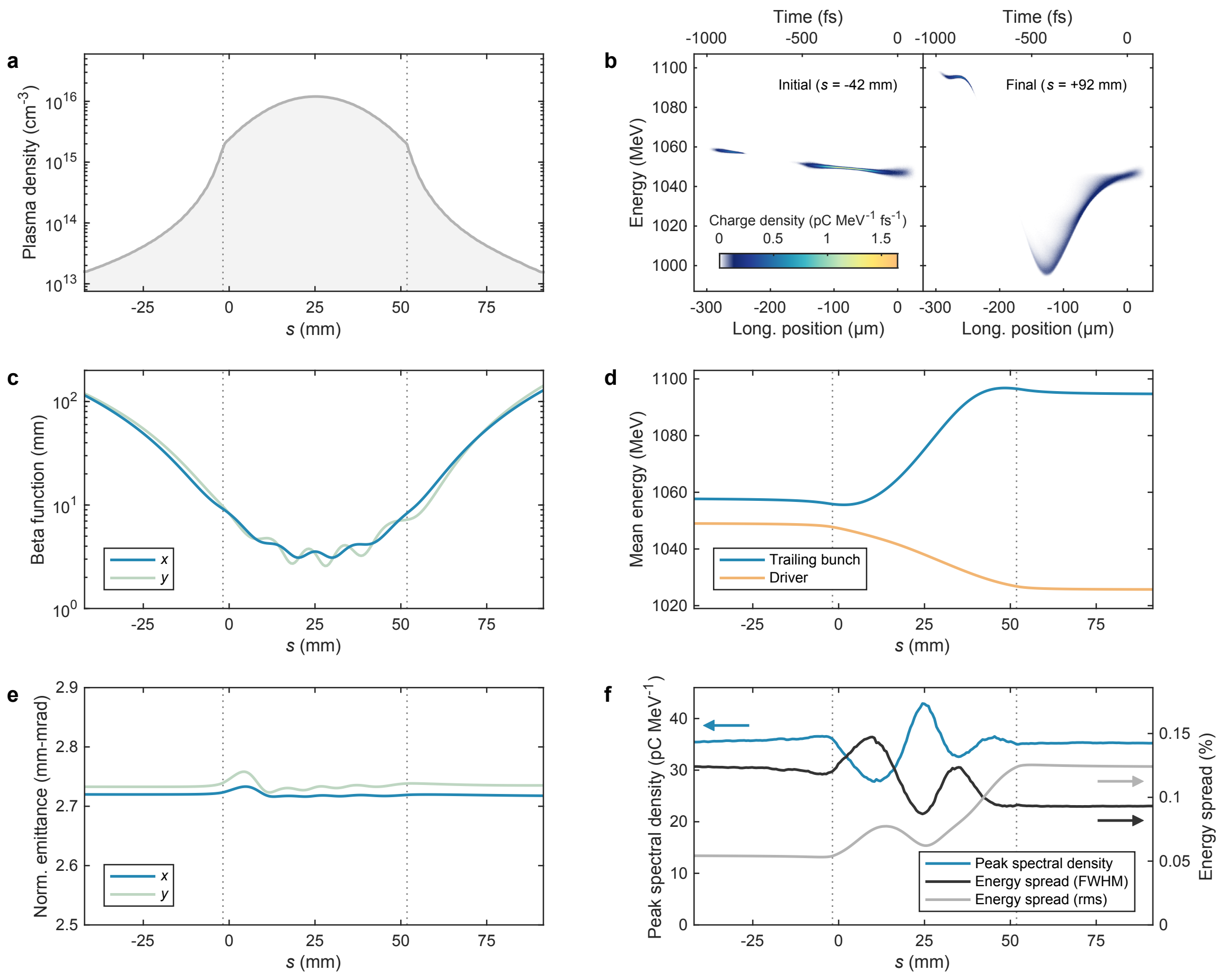

**Supplementary Fig. 10 | Particle-in-cell simulation showing the evolution of the trailing-bunch properties.** **a**, The longitudinal plasma-density profile, shown on a logarithmic scale, was estimated from measurements (Supplementary Fig. 4). Dotted vertical lines indicate the ends of the plasma cell, including its electrodes. **b**, Combined with the measured longitudinal phase space (Supplementary Fig. 3), reproduced in the left panel, the simulation resulted in the longitudinal phase space shown in the right panel. **c**, Between the start and the end of the simulation, the transverse phase space of the trailing bunch also evolved significantly, as indicated by the beta function, shown on a logarithmic scale. The beta functions in both the horizontal ($x$) and vertical ($y$) planes were focused to close to the matched beta function of the plasma accelerator: approximately 3 mm. **d**, Inside the plasma cell, the trailing bunch gained 37 MeV of energy per particle, whereas the driver lost approximately 23 MeV, on average. This resulted in an energy-transfer efficiency of 20%, in agreement with measurements (Supplementary Fig. 6). **e**, As in the experiment, the normalized emittance of the trailing bunch was preserved in the horizontal plane (to within 0.1%). In the simulation, the emittance in the vertical plane was also preserved (also to within 0.1%); however, this quantity was not measured in the experiment. **f**, The energy spread of the trailing bunch, while changing during the acceleration process, was reduced from 0.12% to 0.09% FWHM, consistent with the experimental measurement (Fig. 1f). While the peak spectral density does not decrease, the rms energy spread is seen to approximately double due to the introduction of a low-energy tail (as seen in **b**).

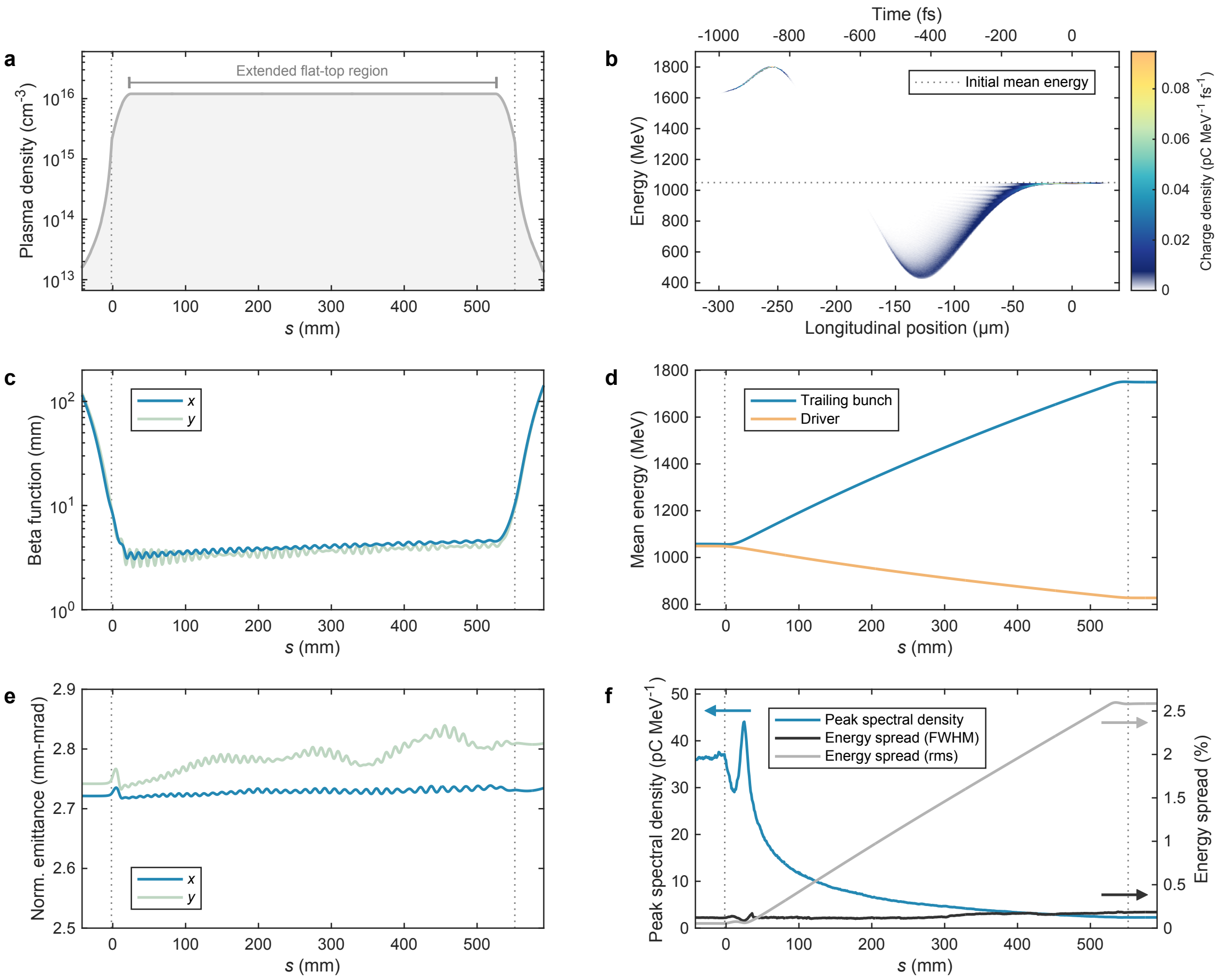

**Supplementary Fig. 11 | Particle-in-cell simulation of an extended plasma cell. a**, The longitudinal plasma-density profile is artificially extended by adding a central 500-mm-long flat-top at $1.2\times10^{16}$ cm$^{-3}$, while keeping the same up and down ramps. All initial beam parameters are identical to that of the simulation in Supplementary Fig. 10. **b**, The resulting longitudinal phase space after acceleration shown, where the dotted line indicates the initial mean energy. **c**, The beta function oscillates with a small amplitude around the matched value, which increases slowly with acceleration. **d**, Here, the trailing bunch gains a total of 692 MeV of energy per particle, whereas the driver loses an average of 222 MeV per particle. In this extended cell, the energy-transfer efficiency is higher (at 40%) than in the shorter cell, as the beam is somewhat over-loaded in the flat-top: an effect that was compensated by under-loading in the low-density ramps in the shorter cell. **e**, Emittance is observed to be approximately preserved, increasing by only 0.5% in the horizontal plane and 2.5% in the vertical plane (both within the measurement error of 3%). Degradation due to hosing and beam-breakup instability is not observed in this simulation. **f**, As a result of the over-loading of the field (seen in **b**), both the FWHM and rms energy spreads increase significantly during acceleration. This could be compensated for by either reducing the beam current of the trailing bunch, or by increasing the amplitude of the wakefield from the driver.